\documentclass[pra,reprint,superscriptaddress,longbibliography]{revtex4-2}

\usepackage{array}
\usepackage{natbib}
\usepackage[section]{placeins}
\usepackage{pgfplots,amsfonts}
\usepackage[breaklinks]{hyperref}
\usepackage{amssymb}
\usepackage{amsmath,amsthm,bm}
\usepackage{amsfonts}
\usepackage{cleveref}
\usepackage{subcaption}
\usepackage{mathtools}
\usepackage{tikz}
\hypersetup{citecolor=red,colorlinks=true,urlcolor=blue}
\usepackage{algorithm,setspace}
\usepackage{algpseudocode}
\usetikzlibrary{angles, quotes}
\usetikzlibrary{positioning}

\usepackage[normalem]{ulem}
\usepackage{color}
\usepackage{xcolor}

\newcommand{\bd}[1]{\boldsymbol{#1}}

\newcolumntype{P}[1]{>{\centering\arraybackslash}p{#1}}
\def\algbackskip{\hskip\dimexpr-\algorithmicindent+\labelsep}
\def\LState{\State \algbackskip}%
\newcommand{\Tr}{\mbox{Tr}}

\newcommand{\bra}[1]{\mbox{$\langle #1 |$}}
\newcommand{\ket}[1]{\mbox{$| #1 \rangle$}}

\definecolor{Ugreen}{HTML}{198a11}

\begin{document}
\title{Combining the contracted quantum eigensolver with the Rayleigh-Ritz variational principle for mixed states for the computation of quantum excited states}

\title{Quantum simulation of excited states from parallel contracted quantum eigensolvers}

\author{Carlos L. Benavides-Riveros }
\email{cl.benavidesriveros@unitn.it}
\affiliation{Pitaevskii BEC Center, CNR-INO and Dipartimento di Fisica, Università di Trento, I-38123 Trento, Italy}

\author{Yuchen Wang}
\affiliation{Department of Chemistry and The James Franck Institute, The University of Chicago, Chicago, Illinois 60637, USA}
\author{Samuel Warren}
\affiliation{Department of Chemistry and The James Franck Institute, The University of Chicago, Chicago, Illinois 60637, USA}

\author{David A. Mazziotti}
\email{damazz@uchicago.edu}
\affiliation{Department of Chemistry and The James Franck Institute, The University of Chicago, Chicago, Illinois 60637, USA}

\date{Submitted November 8, 2023}

\begin{abstract}
Computing excited-state properties of molecules and solids is considered one of the most important near-term applications of quantum computers. While many of the current excited-state quantum algorithms differ in circuit architecture, specific exploitation of quantum advantage, or result quality, one common feature is their rooting in the Schr\"odinger equation. However, through contracting (or projecting) the eigenvalue equation, more efficient strategies can be designed for near-term quantum devices. Here we demonstrate that when combined with the Rayleigh-Ritz variational principle for mixed quantum states, the ground-state contracted quantum eigensolver (CQE) can be generalized to compute any number of quantum eigenstates simultaneously. We introduce two \textit{excited-state} (anti-Hermitian) CQEs that perform the excited-state calculation while inheriting many of the remarkable features  of the original ground-state version of the algorithm, such as its scalability. To showcase our approach, we study several model and chemical Hamiltonians and investigate the performance of different implementations.
\end{abstract}

\maketitle

\section{Introduction}

Calculating physical properties of ex\-ci\-ted-state processes of quantum many-body systems is one of the most promising applications of near-term quantum computing \cite{Cerezo2021,Bauman2021,PhysRevX.6.031007}. Quantum devices are well suited to deal with many of the distinctive features of excited states such as their strong mul\-ti\-con\-fi\-gu\-ra\-tional character or the presence of conical intersections  \cite{doi:10.1021/acs.jpca.6b04932,PhysRevX.8.031022}. So far, several quantum algorithms have been developed to approximate eigenstates of many-body Hamiltonians, including quantum pha\-se estimation (QPE) \cite{PhysRevLett.83.5162,Aspuru-Guzik2005} and the varia\-tio\-nal quantum  eigen\-sol\-ver (VQE) \cite{Peruzzo2014, McClean_2016}. VQE has also inspired several related approaches for excited states: The two dominant variants rely on either targeting specific states through adding nonorthogonal penalties to the Hamiltonian \cite{Higgott2019variationalquantum,PhysRevA.99.062304,PhysRevResearch.4.013173,Wen2021,Shirai} or by building subspaces while ensuring orthogonality of the lowest-lying eigenstates \cite{PhysRevResearch.1.033062,Yalouz_2021}. Yet, QPE requires circuit depths beyond what is currently achievable, and VQE relies on high-dimensional classical optimization, which has computational costs that scale rapidly with the system size \cite{TILLY20221}. 

Quantum algorithms like QPE and VQE are designed to solve the Schr\"odinger equation (SE). However, more efficient quantum simulations can be performed if, instead of the standard SE, its contraction (or projection) is solved directly on a quantum computer \cite{PhysRevLett.126.070504}. When solving the corresponding contracted Schr\"odinger equation (CSE) the prepared wave function ansatz only requires two-body terms, regardless of the number of electrons or orbitals, ensuring the scalability of the algorithm \cite{PhysRevA.69.012507}. While initially designed to explore ground states of molecular systems \cite{PhysRevA.69.012507}, quantum eigensolvers based on the CSE have been recently extended to excited states by using the variance of the energy as the cost function \cite{wang2023electronic} or by deflating the CSE to ensure the orthogonality of the eigenstates \cite{smart2023manybody}. However, these methods compute the eigenstates individually and therefore the circuit must be run for each desired excited state.  

The goal of this work is to demonstrate that when com\-bined with the Rayleigh-Ritz variational principle for mixed quantum states, the CSE can be straightforwardly generalized for the simultaneous (or parallel) calculation of a bundle of lowest eigenstates. Our main result is a novel excited-state quantum algorithm that employs the main features of the ground-state contracted quantum eigensolver (CQE), thus retaining its favorable scaling. Here we focus on the anti-Hermitian portion of the CSE which has been shown to render accurate approximations for ground-state calculations \cite{PhysRevA.75.022505}, but our results can be generalized to include its Hermitian part. In the same way, we focus on fermionic systems but our derivations equally hold for bosons.

The remainder of this paper is organized as follows: For completeness, we first introduce both the CSE and the Rayleigh-Ritz variational principle for ensembles, on which our algorithm is based. Next, we generalize the basic equations of the ground-state CQE to excited states and discuss the resulting quantum algorithm. We then present our contracted quantum eigensolvers, discuss different methods of implementing them on a quantum computer, and perform several numerical experiments. The paper ends with some conclusions and a discussion about potential future research directions.  

\section{Theory}

After we review the CSE and the the Rayleigh-Ritz variational principle for mixed states in sections~\ref{sec:cse} and~\ref{sec:var}, we derive an anti-Hermitian CSE (ACSE) for mixed states in section~\ref{sec:acse} and a quantum algorithm based on this mixed-state ACSE in section~\ref{sec:cqe}, which can solve for multiple excited states simultaneously.     

\subsection{Contracted Schr\"odinger equation}

\label{sec:cse}

The  SE of an electronic system governed by a Hamiltonian $\hat H$ reads:
\begin{align}
(\hat H - E_\nu) \ket{\psi_\nu} = 0.    
\label{SE}
\end{align}
The two-body operator ${\hat \Gamma}^{pq}_{st} \equiv {\hat f}^\dagger_p {\hat f}^\dagger_q{\hat f}_t {\hat f}_s$, where ${\hat f}^\dagger_p / {\hat f}_p$ are fermionic creation/annihilation operators, followed by the vector $\bra{\psi_\nu}$, can be applied on the left of the SE in Eq.~\eqref{SE} to obtain the  CSE:
\begin{align}
  \bra{\psi_\nu} {\hat \Gamma}^{pq}_{st}  (\hat H -E_\nu)\ket{\psi_\nu} = 0.
  \label{CSE}
\end{align}
Both the CSE in Eq.~\eqref{CSE} and the SE in Eq.~\eqref{SE} have an \textit{equivalent} set of pure-state solutions \cite{PhysRevA.14.41, PhysRevA.57.4219,ch8}:  while the SE clearly implies the CSE, the opposite direction is provable by showing that \eqref{CSE} implies the eigenstate condition of zero variance (i.e., $ \bra{\psi_\nu} (\hat H - E_\nu)^2  \ket{\psi_\nu} = 0$) which in turn implies the SE.
Notice that  Eq.~\eqref{CSE} can be written as the sum of two terms (a commutator and anti-commutator) \cite{PhysRevLett.97.143002,Valdemoro2007}:
\begin{align}
  \bra{\psi_\nu} \{{\hat \Gamma}^{pq}_{st},  (\hat H - E_\nu)\} \ket{\psi_\nu} +   \bra{\psi_\nu} [{\hat \Gamma}^{pq}_{st},  \hat H ] \ket{\psi_\nu} = 0\,.
  \label{eq0b}
 \end{align} 
It is well-known that solving only the anti-Her\-mi\-tian portion of this equation, i.e., 
\begin{align}
\bra{\psi_\nu} [{\hat \Gamma}^{pq}_{st} , \hat H ] \ket{\psi_\nu}=0
\label{anti}
\end{align}
gives accurate results both for ground bosonic \cite{wang2023boson} and ground and excited electronic \cite{smart2023manybody, wang2023electronic} states. Moreover, since the Eq.~\eqref{anti} can be interpreted as the residual of a certain cost function, this anti-Hermitian CSE (ACSE) immediately suggests the type of ansatz that can be used to guess the form of the eigenstate $\ket{\psi_\nu}$ (see below). 


\subsection{Variational principle for ensembles}

\label{sec:var}

The Rayleigh-Ritz variational principle is a powerful tool routinely used to study eigenstates  of quantum many-body systems \cite{MSM_1931__49__1_0}. Its generalization to mixed quantum states establishes an upper bound for the weighted ensemble energy of the $K$ lowest eigenstates of a Hamiltonian, $\hat H$  \cite{PhysRevA.37.2809}:
\begin{align}
\Tr\big[\rho(\bd{w})\hat H\big] \geq \sum^{K-1}_{\nu=0} w_\nu E_\nu ,
\label{costSSVQE}
\end{align}
where $\rho(\bd{w}) = \sum^{K-1}_{\nu=0} w_\nu \ket{\phi_\nu}\bra{\phi_\nu}$ is a density matrix with a positive, decreasingly ordered spectrum, conveniently defined as $\bm{w} = (w_0,w_1,\ldots)$ with $w_\nu \geq w_{\nu+1}\geq 0$. The vectors $\{\ket{\phi_\nu}\}$ can be any set of $K$ orthogonal states. Here $E_\nu \leq E_{\nu+1}$ are the exact eigenenergies of the system, arranged in increasing order. The ensemble variational principle in Eq.~\eqref{costSSVQE} offers a unified approach to variational methods in quantum mechanics: the problem of the ground state is, in fact, just a particular case, corresponding to  $\bd{w} = (1,0,0,\ldots)$. This variational approach to quantum excitations is currently playing a pivotal role in the extension of ground-state functional theories \cite{PhysRevLett.124.243001,PhysRevLett.127.023001,PhysRevLett.130.106401,Cernatic2021} and hybrid quantum-classical methods  \cite{PhysRevResearch.1.033062,Yalouz_2021,PhysRevA.107.052423} to excited states.

We note in passing that Eq.~\eqref{costSSVQE} can be written in a state-specific form by employing the purified state \cite{PhysRevLett.129.066401}:
\begin{align}
\ket{\rho(\bd{w})} = \sum_{\nu=0}^{K-1} \sqrt{w_\nu} \,\ket{\phi_\nu} \otimes \ket{a_\nu} \,.
\label{pure}
\end{align}
The states $ \ket{a_\nu}$ are auxiliary orthonormal (ancilla) states added to perform the purification. The only condition is their orthornormality, $\bra{a_\nu}a_\mu\rangle = \delta_{\nu\mu}$. Then, the lower bound of the energy expectation value of the ensemble energy  can be written as  
$\bra{\rho(\bd{w})}\hat H\otimes \mathbb{I}\ket{\rho(\bd{w})} \geq \bd{w}\cdot\bd{E}$, with  $\bm{E} = 
 (E_0,E_1,\ldots)$ and $\mathbb{I}$ being the identity matrix acting on the auxiliary space (we will skip the writing of $\mathbb{I}$ when the notation is obvious).


\subsection{The ACSE for excited states}

\label{sec:acse}

The generalization of the CSE to ensembles of eigenstates is straightforward. Indeed,  since Eq.~\eqref{CSE}  is valid for all the eigenstates of the Hamiltonian $\hat H$, one can use it  to write a weighted sum for the first $K$ eigenstates:
$\sum^{K-1}_{\nu = 0} w_\nu \bra{\psi_\nu} {\hat \Gamma}^{pq}_{st}  (\hat H - E_\nu) \ket{\psi_\nu} = 0$. From this equation, the corresponding ACSE for an ensemble of $K$ eigenstates follows:
\begin{align}
  \sum^{K-1}_{\nu = 0} w_\nu \bra{\psi_\nu} [{\hat \Gamma}^{pq}_{st} , \hat H ] \ket{\psi_\nu} = 0\,.
    \label{eq0e}
\end{align}
This result suggests a variational implementation of the ACSE for excited states. Consider first a variational ansatz for a set of $K$ orthogonal wave functions, iteratively constructed from unitary two-body exponential transformations:
\begin{align}
  \ket{\phi^{(n+1)}_\nu} = e^{\eta \hat A^{(n)}}   \ket{\phi^{(n)}_\nu}\,,
  \label{eACSE}
\end{align}
where 
$\hat A^{(n)} = \sum_{pq,st} A^{(n)}_{pq,st}  {\hat f}^\dagger_p {\hat f}^\dagger_q{\hat f}_t {\hat f}_s$ is an anti-Hermitian two-electron operator and $\eta$ is a real positive number (who\-se role will be clear later). The ensemble energy at the $(n+1)$th iteration is the weighted sum of the energy ex\-pectation value of these states:
\begin{align}
\mathcal{E}_{n+1} \equiv  \sum^{K-1}_{\nu = 0} w_\nu E_\nu^{(n+1)} =  \sum^{K-1}_{\nu = 0} w_\nu\bra{\phi_\nu^{(n+1)}}  \hat H \ket{\phi_\nu^{(n+1)}}\,,
\end{align}
Thus, at each iteration, the ensemble energy through order $\eta$ is 
$\mathcal{E}_{n+1} = \mathcal{E}_{n}  + \eta \sum_\nu w_\nu \bra{\phi^{(n)}_\nu} [\hat H, \hat A^{(n)}]  \ket{\phi^{(n)}_\nu} + \mathcal{O}(\eta^2)$. As in the case of the ground-state calculation \cite{PhysRevLett.126.070504}, the gradient of the ensemble energy can be computed with respect to each $A^{(n)}_{pq,st}$:
\begin{align}
    \frac{\partial \mathcal{E}_{n}}{\partial A^{(n)}_{pq,st}} = \eta  \sum_\nu w_\nu r^{(n)}_{\nu;pq,st}\,.
    \label{residual}
\end{align}
where  $r^{(n)}_{\nu;pq,st} \equiv \bra{\phi^{(n)}_\nu}[\hat H, {\hat \Gamma}^{pq}_{st}] \ket{\phi^{(n)}_\nu}$. This shows that the residual of the energy is the \textit{weighted} expectation value of the commutators $[\hat H, {\hat \Gamma}^{pq}_{st}]$. The residual goes to zero when the ensemble is composed of eigenstates, which means that the ACSE in Eq.~\eqref{eq0e} is fulfilled. Hence, an algorithm to find the optimal operator $\hat{A}$ using gradient descent should perform the following update of the parameters at each step:
\begin{align}
     A^{(n+1)}_{pq,st} =  A^{(n)}_{pq,st} -   \frac{\partial \mathcal{E}_{n}}{\partial A^{(n)}_{pq,st}} \,,
\end{align}
which implies that $\eta$ is the learning rate of the algorithm.

Interestingly, the purification introduced in Eq.~\eqref{pure} can be used to write a more compact expression for the residual of the \textit{ensemble} ACSE in Eq.~\eqref{residual}, namely: $\bra{\rho(\bd{w})} [{\hat \Gamma}^{pq}_{st}, \hat H ]\otimes \mathbb{I} \ket{\rho(\bd{w})}$. If, in addition, one chooses the auxiliary states as a replica of the physical ones (i.e., $\ket{a_\nu} = \ket{\phi_\nu}$), then the state can be written as the following unitary transformation of the system's vacuum \cite{PhysRevLett.129.066401}: $\ket{\rho(\bd{w})}=V(\bd{w})\ket{0}$, where $V(\bd{w})= UD(\bd{w})$, $U$ is a unitary acting on the physical space and $D(\bd{w})$ is a squeezed operator acting on the duplicate Hilbert space. As a result, the total residual can be written as a vacuum expectation value:
$\bra{0} [{\hat \Gamma}^{pq}_{st}(\bd{w}), \hat H(\bd{w}) ] \ket{0}$, where the notation $\hat A(\bd{w}) = V^\dagger(\bd{w}) \hat A V(\bd{w})$  is used. 
\begin{figure}[!t]
\begin{algorithm}[H]
\begin{spacing}{1}
\setstretch{1.09}
\begin{algorithmic}[1]
\State Given $K>0$, $\bd{w}= (w_0,..,w_{K-1})$, $\delta > 0$, 
\State choose $0 < \eta < 1$,
\State choose $K$ physical and ancilla states $\{\ket{\phi_\nu},\ket{a_\nu}\}^{K-1}_{\nu = 0}$,
\State initialize the state $\ket{\rho_{0}(\bd{w})} = \sum_{\nu=0}^{K-1} \sqrt{w_\nu} \ket{\phi_\nu}\otimes \ket{a_\nu}$.
\State Set $n \gets 0$,
\While{$|A^{(n)}|^2 > \delta$}
\State prepare the state $\ket{\Lambda^\pm_n} = e^{\pm i \eta \hat H} \ket{\rho_n(\bd{w})}$,
\State measure $A^{(n)}_{pq,st} = \frac{1}{2i}[\bra{\Lambda^+_n} {\hat \Gamma}^{pq}_{st}\ket{\Lambda^+_n}-\bra{\Lambda^-_n} {\hat \Gamma}^{pq}_{st}\ket{\Lambda^-_n}]$,
\State prepare $\ket{\Sigma_n(\theta)} = \exp(\theta \hat A^{(n)}) \ket{\rho_n(\bd{w})}$,
\State minimize $\bra{\Sigma_n(\theta)}\hat H\ket{\Sigma_n(\theta)}$ with respect to  $\theta$,
\State take $\theta^* = {\rm argmin} \bra{\Sigma_n(\theta)}\hat H\ket{\Sigma_n(\theta)}$,
\State prepare $\ket{\rho_{n+1}(\bd{w})} = \exp(\theta^* \hat A^{(n)}) \ket{\rho_n(\bd{w})}$,
\State $n \gets n+1$.
\EndWhile
\end{algorithmic}
\caption{Parallelized CQE}
\label{alg:cap}
\end{spacing}
\end{algorithm}
\end{figure}

One possible way to implement the ACSE in a quantum device is to choose $w_\nu$ as fixed quantities and, for the $(n+1)$th iteration, allocate a certain number of shots $N_\nu$ to measure the contribution of $r_{\nu;pq,st}$ to the total residual in Eq.~\eqref{residual}. Yet it is known that the most efficient way of deterministic assigning shots among the measurements consists of allocating $N_\nu$ proportionally to $w_\nu$ \cite{Rubin_2018,Yen2023}. But since the weights are not integers, this assignment results in a ``hard floor'' on $N_{\rm total} = \sum_\nu N_\nu \geq 1/ w_K$ (recall that $w_K$ is the minimum of the weights) \cite{arrasmith2020operator}. This is the minimal number of shots needed for an unbiased estimate of the residuals of the ensemble $\sum_\nu w_\nu r^{(n)}_{\nu;pq,st}$. Unfortunately, for large $K$ one can expect quite small $w_K$ and therefore very large numbers of shots for each unbiased estimate. Random sampling can efficiently perform unbiased estimations of the residuals of the ensemble energy in Eq.~\eqref{eq0e} while using a cheap number of shots. In the next section, based on this sampling, we will present two quantum algorithms. 

\subsection{CQE for excited states}

\label{sec:cqe}

\begin{figure}[!t]
\begin{algorithm}[H]
\begin{spacing}{1}
\setstretch{1.09}
\begin{algorithmic}[1]
\State Given $K>0$, $\bd{w}= (w_0,..,w_{K-1})$, $\sum_\nu w_\nu = 1$, $\delta > 0$, 
\State choose $0 < \eta < 1$, and $N >0$ number of shots,
\State choose $K$ initial states $\{\ket{\phi^{(0)}_0},...,\ket{\phi^{(0)}_{K-1}}\}$,
\State Set $n \gets 0$,
\While{$|A^{(n)}|^2 > \delta$}
\State $\bd{m} \sim {\rm Multinomial}(N, \bd{w})$
\State Set $A^{(n)}_{pq,st} \gets 0$ 
 \For {$0\leq \nu \leq K-1$}
 \For {$1\leq l \leq m_\nu$}
        \LState prepare $\ket{\lambda^\pm_{\nu}} =  e^{\pm i \eta \hat H} \ket{\phi^{(n)}_\nu}$,
        \LState $A^{(n)}_{pq,st} \gets A^{(n)}_{pq,st} +  \frac{1}{2i}\sum_{z=\pm} z \bra{\lambda^z_\nu} {\hat \Gamma}^{pq}_{st}\ket{\lambda^z_\nu}$,
      \EndFor
      \State prepare $\ket{\Sigma_\nu(\theta)} = e^{\theta \hat A^{(n)}} \ket{\phi_\nu^{n}}$,
      \EndFor
\State take $\theta^* = {\rm argmin} \sum_\nu w_\nu\bra{\Sigma_\nu(\theta)}\hat H\ket{\Sigma_\nu(\theta)}$,
\State prepare $\ket{\phi^{(n+1)}_\nu} = \exp(\theta^* \hat A^{(n)}) \ket{\phi^{(n)}_\nu}$,
\State $n \gets n+1$.
\EndWhile
\end{algorithmic}
\caption{Weighted random CQE}
\label{alg:cap2}
\end{spacing}
\end{algorithm}
\end{figure}

To introduce our algorithms, let us start first by choosing a set of weights $\bd{w}$, which for convenience we normalize to 1: $\sum_\nu w_\nu =1$. Next, we choose $K$ initial orthogonal states $\ket{\phi_\nu}$ that can be the $K$ lowest mean-field (Hartree-Fock) wave functions. Weighted random sampling, where the probability of measuring $r_{\nu}$ is proportional to $w_\nu$, can be used as an efficient unbiased estimator of the residuals of the ensemble energy in Eq.~\eqref{eq0e}. A promising alternative to implementing this anti-Hermitian CQE that does not require a random number generator consists of preparing and measuring the purification presented in Eq.~\eqref{pure}. For this \textit{parallelized CQE} the initial state in Eq.~\eqref{pure} can be prepared by applying a suitable linear combination of unitaries \cite{Childs2012} to the original Har\-tree-Fock state 
 $\ket{\rho_0}=\ket{\phi_{\rm HF}}\otimes\ket{0,...,0}$,
 with an ancilla term that uses only $\log_2K$ qubits. 
 At each iteration, the states $\ket{\Lambda^\pm_n} = \exp(\pm i \eta \hat H) \ket{\rho_n(\bd{w})}$ are prepared and the entries of the matrix $A^{(n)}$ are measured from the equation 
 \begin{align*}
 A^{(n)}_{pq,st} = \frac{1}{2i}(\bra{\Lambda^+_n} {\hat \Gamma}^{pq}_{st}\ket{\Lambda^+_n}-\bra{\Lambda^-_n} {\hat \Gamma}^{pq}_{st}\ket{\Lambda^-_n}) + \mathcal{O}(\eta^2).
 \end{align*}
 Importantly,  the residual in Eq.~\eqref{residual} is exactly zero for any set of eigenstates, not necessarily the lowest ones, so for any combination of eigenstates the optimization will stop at this point. Hence, to guarantee that the lowest set is found, we further prepare the state $\ket{\Sigma_{n}(\bd{w})} = \exp(\theta \hat A^{(n)}) \ket{\rho_n(\bd{w})}$ and minimize the ensemble energy with respect to the value of $\theta$. Besides circumventing local minima,  this will also guarantee a faster convergence. As described in Algorithm \ref{alg:cap}, the process is repeated until a desired convergence is reached.

 We also sketch the \textit{weighted random CQE} in Algorithm \ref{alg:cap2}. This algorithm follows similar lines as Algo\-rithm \ref{alg:cap} except that the purification (or the paralleli\-zation) is replaced by assigning $m_\nu$ number of shots per state $\ket{\phi_\nu}$ randomly from a multinomial distribution: $\bd{m} \sim {\rm Multinomial}(N_{\rm total}, \bd{w})$.  Because each of the excited states is treated separately, the algorithm is amenable to distributed parallel programming in which each state is prepared and measured on a separate quantum processor with the results only collected for the classical parts of the optimization.  This weighted random sampling algorithm is \textit{equivalent} to measuring the expectation value of the pure state $\ket{\rho(\bd{w})}$, in the sense that the variance of any observable computed by both methods does coincide. 
As a result, the number of shots needed to achieve a certain measurement error of the residuals is the same for both algorithms. Yet, while the results are certainly the same, the implementation clearly differs in the requirement of computational resources. An advantage, however, of the purification lies in the fact that quantum symmetries can easily be added to the cost function to improve convergence \cite{PhysRevB.94.041116,Lyu2023symmetryenhanced}.  

\section{Results}

\begin{figure}[!t]
  \begin{tikzpicture}
 \node (img) {\includegraphics[scale=0.35]{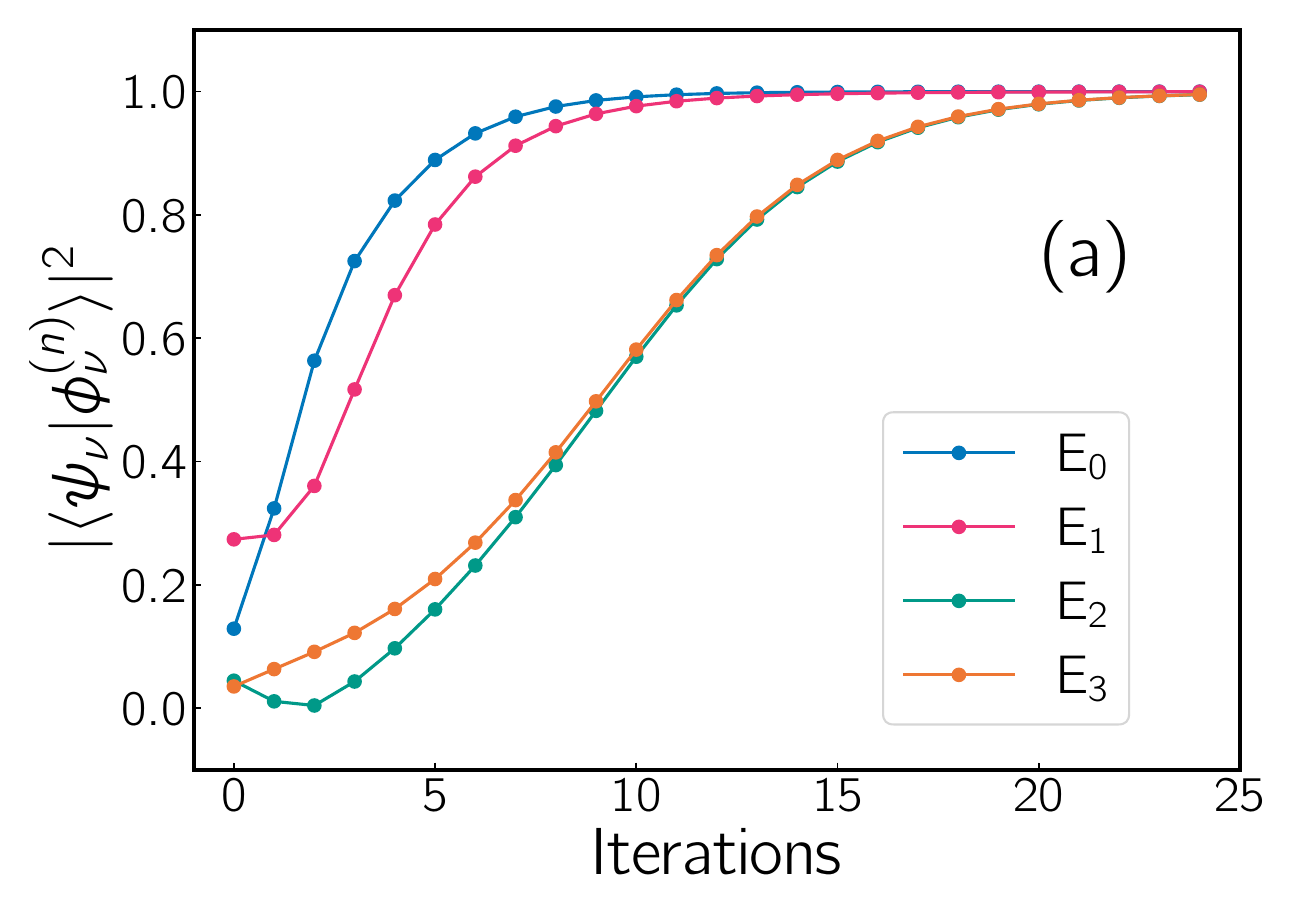}};
 \node[left=of img, node distance=0cm, anchor=center, xshift=2.1cm,yshift=2.7cm,font=\color{black}] {};
 \end{tikzpicture}
  \begin{tikzpicture}
 \node (img) {
  \includegraphics[scale=0.35]{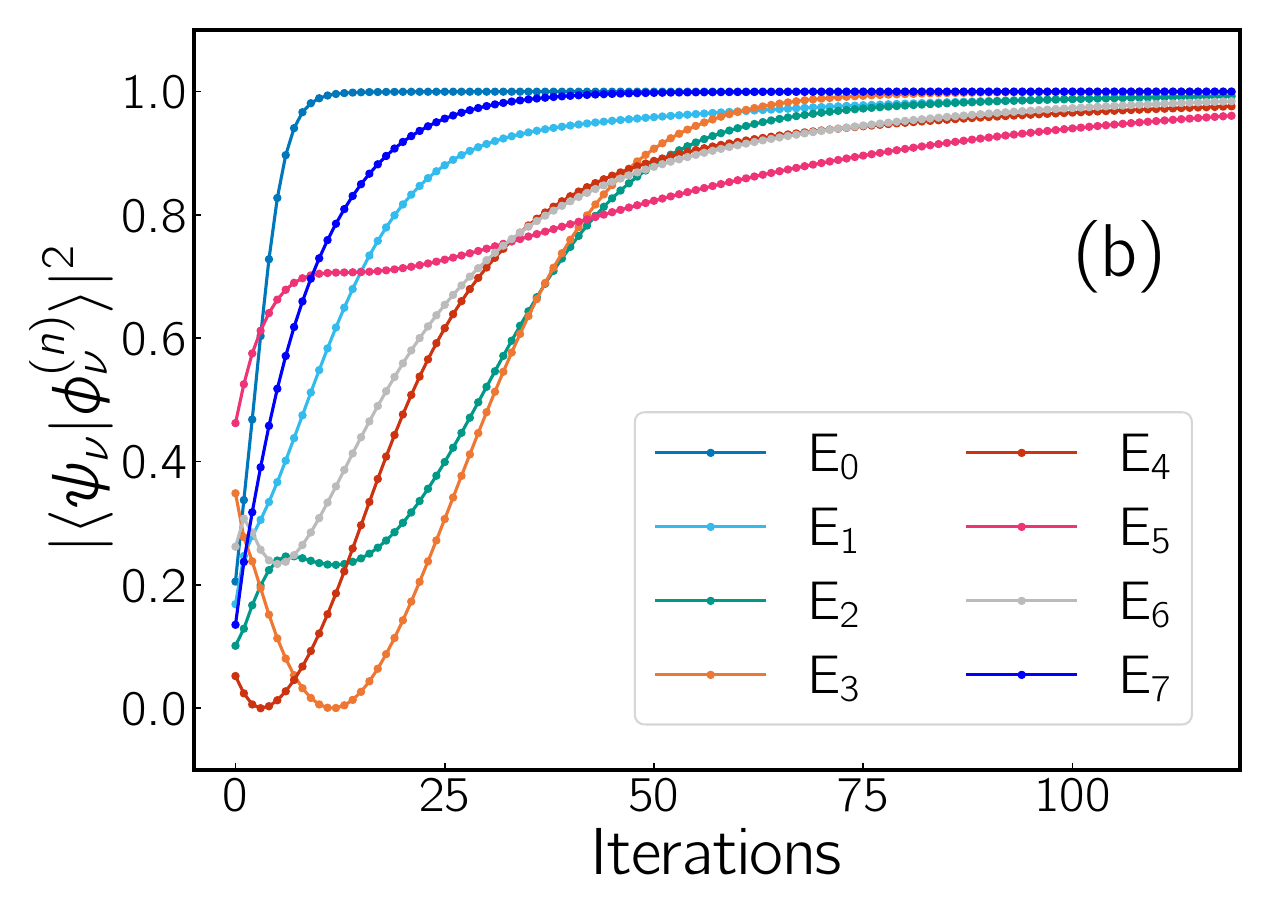}};
   \node[left=of img, node distance=0cm, anchor=center, xshift=2.1cm,yshift=2.7cm,font=\color{black}] {};
 \end{tikzpicture}
\caption{Evolution of the projection of the states $\ket{\phi_\nu^{(n)}}$ on the exact eigenstates $\ket{\psi_\nu}$ as a function of the iteration $n$ for (a) 2-qubit and (b) 3-qubit random Hamiltonians in Eq.~\eqref{hamiltonian}.}
\label{fig1}
\end{figure}

We now present the results of both Algorithms when applied to model and molecular Hamiltonians and discuss their advantages and disadvantages. The first system we investigate with the ensemble ACSE is the  generic $M$-qubit Hamiltonian:
\begin{align}
    \hat H = \sum_{r_1,...,r_M} \lambda_{r_1,...,r_M} \bigotimes_{n=1}^M\sigma_n  \,,
    \label{hamiltonian}
\end{align}
where $\sigma_n$ denotes the Pauli matrix. The initial state is denoted as $\ket{\rho_0(\bd{w})} = \sum_{\bd{i}\in \{0,1\}^M}\sqrt{w_{\bd{i}}} \, \ket{\bd{i}}_p\otimes \ket{\bd{i}}_a$,
whe\-re $p/a$ denotes the physical/ancilla qubits and $\bd{i}=(i_1,...,i_M)$.  The evolution into the exact eigenstates for a random Hamiltonian of the form in Eq.~\eqref{hamiltonian} for systems sizes $M=2,3$  is presented in Fig.~\ref{fig1}. We chose the learning rate $\eta = 0.3$ and weights $\bd{w} = (M^2,M^2-1,...,1)$ and then $\bd{w} \rightarrow \bd{w}/\sum_i w_i$. For $M=2$, the ground state is reached in 8 iterations, while the exact eigenstate calculation is reached in 20. The highest energy states, having the lowest weights in the cost function, converge the slowest, and, due to orthogonality limiting the degrees of freedom, converge simultaneously. A similar pattern can be seen for another random Hamiltonian for the case $M=3$ but due to the larger dimension of the Hilbert space, more iterations are needed for convergence.

We investigate also two molecular examples: a noisy backend simulation of H$_{2}$ and a noiseless state-vector simulation of H$_{4}$. All calculations were performed using the minimal Slater-type orbital (STO-3G) basis set. The noisy backend is the FakeLagosV2 by IBMQ. 

 \begin{figure*}[!t]
  \begin{minipage}{0.5\textwidth}
    \centering
    \includegraphics[width=1.0\textwidth]{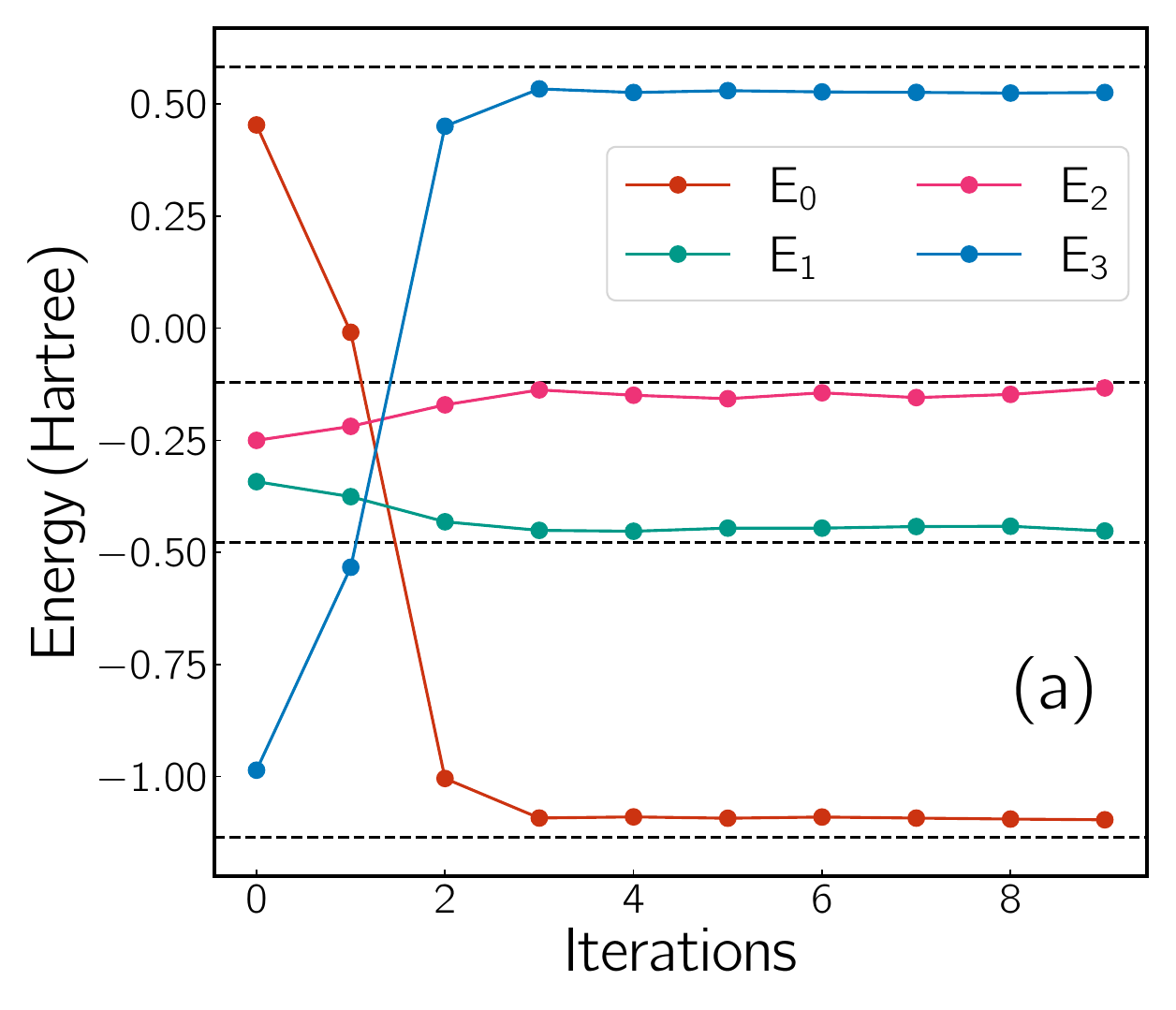}
    \phantomcaption
  \end{minipage}%
  \begin{minipage}{0.5\textwidth}
    \centering
    \includegraphics[width=1.0\textwidth]{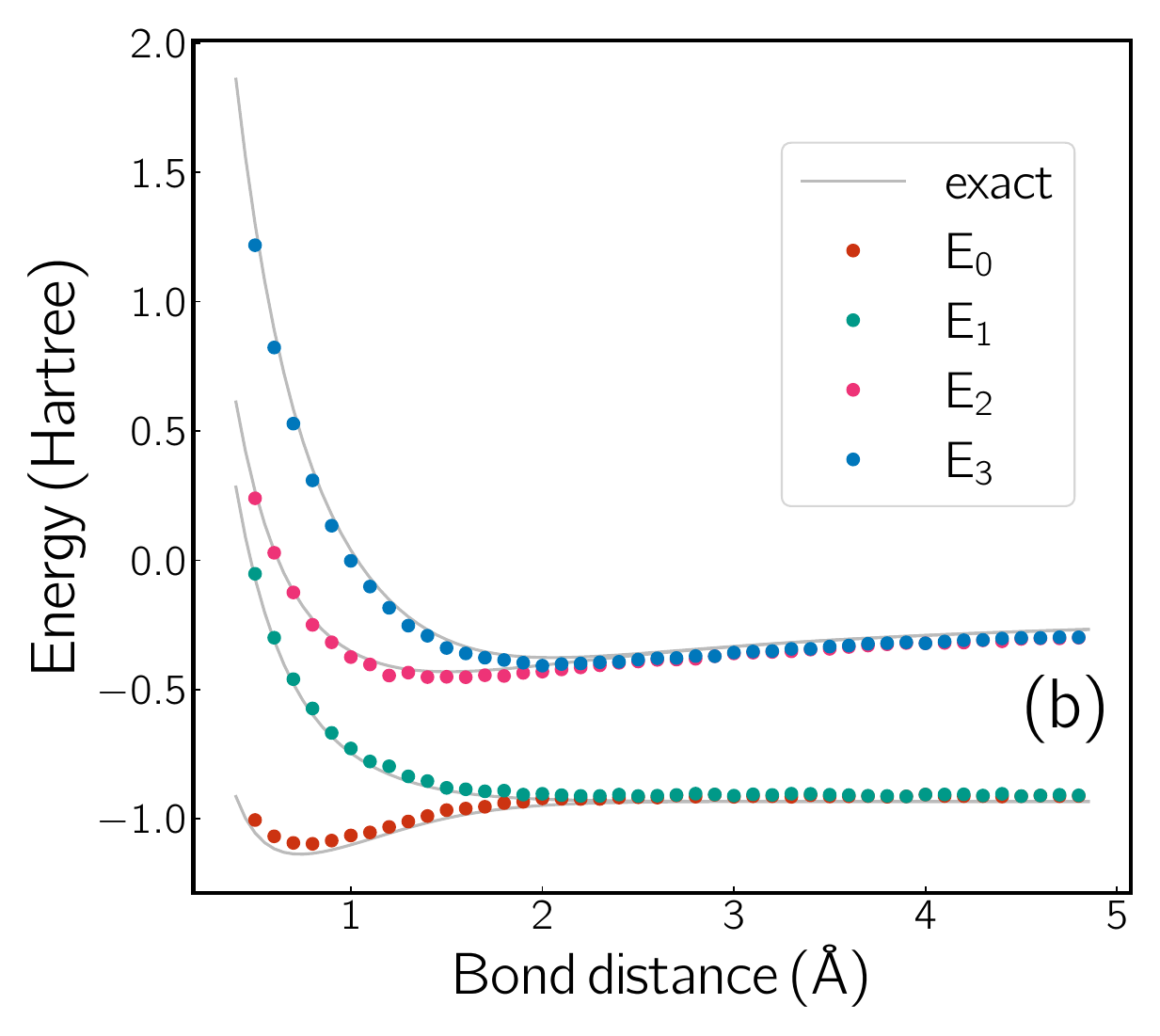}
    \phantomcaption
  \end{minipage}
  \caption{(a) Obtained energies during the optimization for single point calculation of H$_{2}$ (bond distance of 0.7 \AA). The exact solutions for each state and the ensemble are indicated by black dashed lines. (b) Energies along the dissociation curve computed from 0.5 to 5 \AA. The exact results are shown as black lines and our single-point calculations are shown as dots.}
  \label{fig3}
\end{figure*}

Calculation of H$_{2}$ is performed in the spin-symmetry sector $S_{z}=0$. Based on the symmetry of the problem, we construct the Hamiltonian in a compressed form with two qubits. Two additional ancillary qubits are used to create the purified ensemble of all four eigenstates, resulting in four qubits in total. The detailed circuit preparation has been reported in previous work \cite{hong2023quantum}. For the single-point calculation in Fig.~\ref{fig3}, performed with the paralleled CQE, the ensemble energy converges to a minimum in only three iterations.  Remarkably, we achieve an error of less than 30 mHartree for each state without any error mitigation techniques. We also present the dissociation curve of H$_{2}$ in Fig.~\ref{fig3}. Energies computed from parallel CQE are in excellent agreement with the full CI results with an average mean unsigned error of 26 mHartree.

It is also worth discussing the role weight values $w_i$ play in the rate of convergence. For instance, if all of the weights are equal, only an eigen-subspace can be found, and the individual eigenstates would have to be resolved with classical diagonalization. Giving different values for the weights allows us to perform the entire calculation on a quantum device, resulting in a faster convergence. Indeed, we find that the optimal convergence for H$_{2}$ (presented in Fig.~\ref{fig3}) is achieved with the weights $\bd{w} \sim (9,9,1,1)$, before normalization. To explain our choice, let's observe that, due to system's point-group symmetry, the Hamiltonian matrix is block diagonal with two $2\times 2$ sub-matrices on the diagonal. Therefore, since the minimization runs independently within each sub-block, we opted for two identical pairs of weights. This indicates that the optimal choice of weights is highly dependent on the molecular symmetries.

\begin{figure}[t]
\centering
 \includegraphics[scale=0.38]{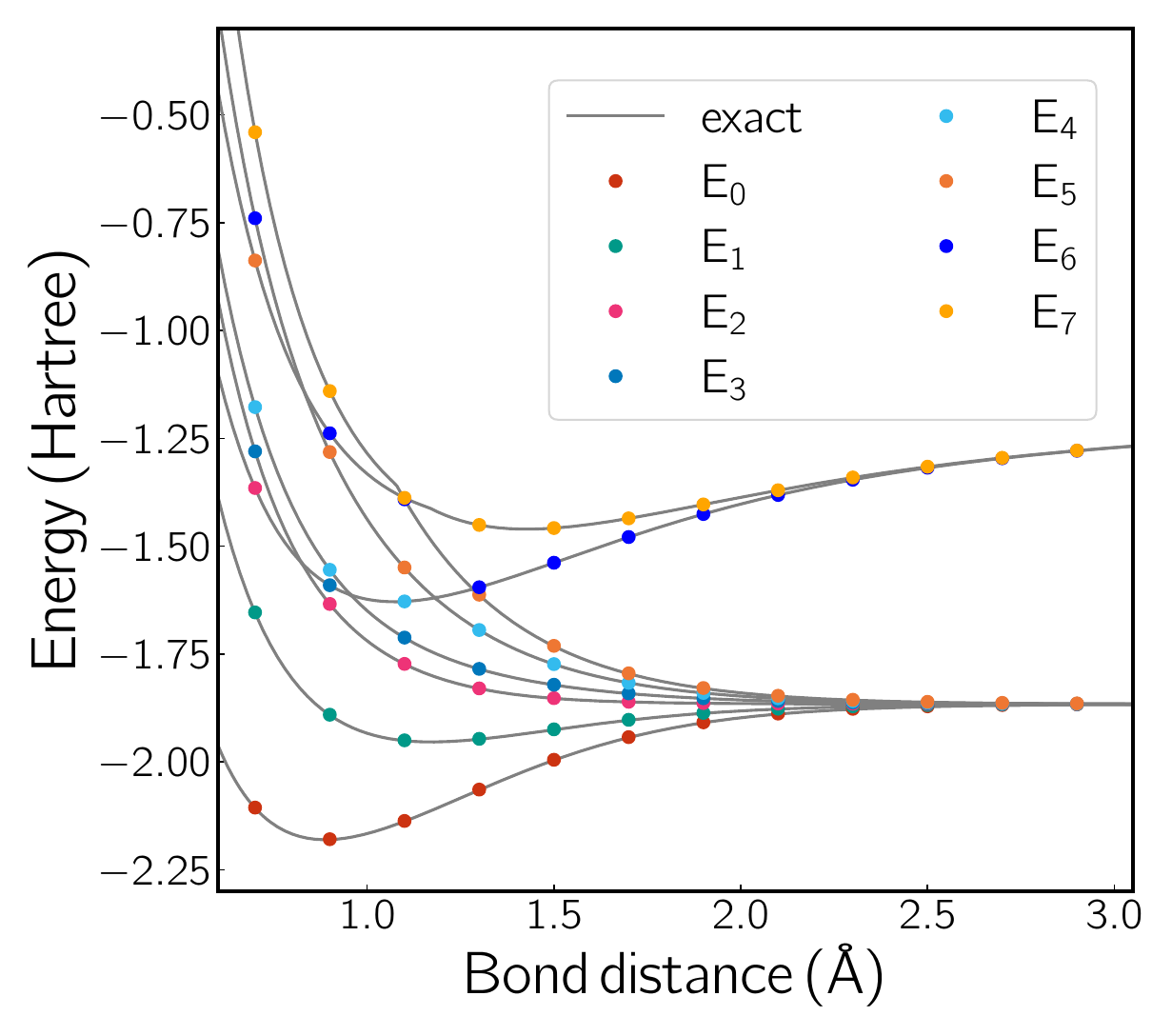}
\caption{The computed and exact lowest eight eigenenergies of the linear equidistant H$_{4}$ as a function of the H-H distance.}
\label{fig2}
\end{figure}

Linear H$_{4}$ is a widely used benchmark system for strong correlation in electronic structure theory \cite{PhysRevResearch.3.013039,Lyu2023symmetryenhanced}. As the molecule dissociates, the energy levels become highly degenerate due to the non-interacting hydrogen atoms and the system exhibits significant static correlation \cite{C7CP01137G}. We take the equidistant form of linear H$_{4}$ and use the Jordan-Wigner transformation to map the Hamiltonian from four spatial orbitals to eight qubits. Both algorithm \ref{alg:cap} and \ref{alg:cap2} successfully find the ground and excited states. Yet in the first case, as we are tackling eight states simultaneously, one requires at least three ancillary qubits to prepare all initial states in the expanded Hilbert space. Alternatively, preparing different initial states separately and sampling them using a multinomial distribution (as in our Algorithm  \ref{alg:cap2}) becomes particularly valuable with limited qubit resources or when the ancilla-based preparations are hard to perform. 

For the calculation of H$_{4}$ shown in Fig.~\ref{fig2}, we have used the weight vector $(8,7,...,1)$, before normalization. At a long bond distance, we seed the eight initial guesses with the eight single Slater determinants with the lowest energies. Afterward, each state in the calculation is seeded with the two most important Slater determinants of the corresponding state found in the previous calculation.  While the potential energy curves are highly degenerate towards dissociation, as the bond begins the form, the energy curves separate.  As shown in Fig.~\ref{fig2}, for the dissociation curve on a noiseless simulator, our algorithms give almost exact results (i.e., an error of around 10$^{-4}$ Hartree). Most calculations converged in less than 200 iterations. We recall that this convergence speed does depend on the weight being assigned to each element, the initial guess, as well as the optimization method, suggesting opportunities for further exploration and improvement. 

\section{Conclusions}

In this paper, we have combined the contracted quantum eigensolver (CQE), originally developed for the calculation of molecular ground states, and the Rayleigh-Ritz variational principle for ensemble states into an excited-state CQE. Quite remarkably, our scheme allows us to compute simultaneously an arbitrary number of lowest eigenstates while preserving the favorable scaling and ease of implementation of the ground-state CQE. Unlike approaches based on the unitary coupled cluster and related ans{\"a}tze, that give an approximation to the cost function, our algorithm provides a natural choice for the unitary operator through the measured residual. In our experiments with molecular and model systems, we tackle multiple states simultaneously with excellent accuracy in both the weakly and strongly correlated regimes. The ability to optimize near-degenerate states by assigning different weights allows us to study both near-degeneracy and conical intersections, which can be used for nonadiabatic chemistry.  Another interesting question for the future is how to use our algorithms for excited states and spectroscopy when additional bosonic degrees are present. 

\textit{Code availability.---} All codes to reproduce, examine, and improve our proposed analysis are freely available online \footnote{\url{https://github.com/damazz/Parallel-CQE}.}. \\

\begin{acknowledgments}
C.L.B.-R. gratefully acknowledge financial support from the European Union’s Horizon Europe Re\-search and Innovation program  un\-der the Marie Skło\-dowska-Curie grant agreement n°101065295.  D.A.M gratefully acknowledges the U.S. Department of Energy, Office of Basic Energy Sciences, Grant DE-SC0019215 and the U.S. National Science Foundation Grant No. CHE-2155082.
\end{acknowledgments}
  
\bibliography{Refs2}
\end{document}